\pdfoutput=1

\documentclass[twocolumn,pra, aps,superscriptaddress,showpacs]{revtex4-1}

\usepackage[colorlinks=true,citecolor=myblue,linkcolor=myblue,urlcolor=myblue]{hyperref}
\usepackage{mathptmx}
\usepackage{subfigure}
\usepackage{dcolumn}
\usepackage{amsmath,amssymb}
\usepackage{bm}
\usepackage{color}
\usepackage{latexsym}
\usepackage{color}
\usepackage[english]{babel}
\usepackage{latexsym}
\usepackage{psfrag,graphicx}
\usepackage{graphicx}
\usepackage{epsf}
\usepackage{amsmath}
\usepackage{amssymb}
\usepackage{amsfonts}
\usepackage{bm}
\usepackage{natbib}
\usepackage{epstopdf}
\usepackage{appendix}
\usepackage[export]{adjustbox}

\definecolor{mygrey}{gray}{0.35}
\definecolor{myblue}{rgb}{0.2,0.2,0.8}
\definecolor{myzard}{cmyk}{0,0,0.05,0}
\definecolor{mywhite}{rgb}{1,1,1}
\definecolor{myred}{rgb}{1,0.,0.3}

\def\be{\begin{equation}}
\def\ee{\end{equation}}
\def\ba{\begin{align}}
\def\enda{\end{align}}
\def\bi{\begin{itemize}}
\def\ei{\end{itemize}}

 \def\ee{\mathord{\rm e}}

\def\min{\mathord{\rm min}}

 \def\ee{\mathord{\rm e}}

\def\min{\mathord{\rm min}}

\renewcommand{\ee}{{\rm e}}

\def\beq{\begin{equation}}
\def\beq{\begin{equation}}
\def\eeq{\end{equation}}

 \newcommand{\ket}[1]{|#1\rangle}
 \newcommand{\bra}[1]{\langle #1|}

 \newcommand{\ketbradif}[2]{\ket{#1}\bra{#2}}
 \newcommand{\ketbra}[1]{\ketbradif {#1}{#1}}

\begin{document}

\title[Short Title]{Fully robust qubit in atomic and molecular three-level systems}

\author{N. Aharon\textsuperscript{1}, I. Cohen\textsuperscript{1}, F. Jelezko\textsuperscript{2,3}, and A. Retzker}

\affiliation{Racah Institute of Physics, The Hebrew University of Jerusalem, Jerusalem
91904, Givat Ram, Israel\\
\textsuperscript{2}Center for Integrated Quantum Science and Technology, Universit\"{a}t Ulm, D-89081 Ulm, Germany\\
 \textsuperscript{3}Institut f\"{u}r Quantenoptik, Albert-Einstein Allee 11, Universit\"{a}t Ulm, D-89081 Ulm, Germany}

\begin{abstract}
{We present a new method of constructing a fully robust qubit in a three-level system. By the application of continuous driving fields, robustness to both external and controller noise is achieved. Specifically, magnetic noise and power fluctuations do not operate within the robust qubit subspace. Whereas all the continuous driving based constructions of such a fully robust qubit considered so far have required at least four levels, we show that in fact only three levels are necessary. This paves the way for simple constructions of a fully robust qubit in many atomic and solid state systems that are controlled by either microwave or optical fields. We focus on the NV-center in diamond and analyze the implementation of the scheme, by utilizing the electronic spin sub-levels of its ground state. In current state-of-the-art experimental setups the scheme leads to improvement of more than two orders of magnitude in coherence time, pushing it towards the lifetime limit. We show how the fully robust qubit can be used to implement quantum sensing, and in particular, the sensing of high frequency signals.}
\end{abstract}
\maketitle

\section{Introduction} The implementation of quantum technology
applications and quantum information processing requires a reliable
realization of qubits that can be initialized, manipulated, and
measured efficiently. In solid state and atomic systems, ambient magnetic
field fluctuations constitute a serious impediment, which usually limits
the coherence time to several orders of magnitude less than the lifetime
limit. Pulsed dynamical decoupling \cite{Han,CP,CPMG} has proven to be very useful in
prolonging the coherence time \cite{Viola,Biercuk,Du,Lange,Ryan,Naydenov,Zhi,BarGil}. However, in order to mitigate both
external and controller noise, very rapid and composite pulse sequences
must be applied \cite{Khodjasteh,Uhrig,Souza,Yang,Farfurnik}, which are not easily incorporated into other operations and require a lot of power \cite{kurizki}.
Similarly, in continuous dynamical decoupling \cite{Fanchini,Bermudez1,Bermudez2,Cai,Xu,Golter,rabl2009,kurizki,Jens2010}, the effect of the controller noise can be diminished by either a rotary echo scheme \cite{Mkhitaryan1,Mkhitaryan2},
which is then analogous to pulsed dynamical decoupling,
or by the concatenation of several driving fields \cite{CaiCon,Itsik1,pathrick},
which is limited by the reduction of the dressed energy gap, and
results in slower qubit gates. However, a multi-state system enables a different approach.
In \cite{Christof}, a fully robust qubit; i.e., a qubit that is robust to both external
and controller noise, was realized by the application of continuous
driving fields on a specific hyperfine structure. Subsequently, a general
scheme for the construction of a fully robust qubit was introduced in \cite{ADR}.

So far, all the continuous driving based implementations of a fully robust
qubit have been investigated \cite{ADR,Itsik2,Wang,Itsik3,Itsik4} and experimentally realized \cite{Christof,Tan,Webster,Randall,Baumgart} with the application
of on-resonance driving fields. This, however, requires at least four
energy levels on which the driving fields operate, and hence is not
applicable to a three-level system. In fact, together with a three-level
system, an additional hyperfine level was considered in \cite{Christof}.
In \cite{Wang}, one of the excited states of the NV-center was
used, but necessitated a cryogenic temperature, and in \cite{ADR}
two $\Lambda$ systems (composed of six states) were employed.

In this paper we show how a fully robust qubit can be constructed by only utilizing
a three-level system through the application of continuous \emph{off-resonant}
driving fields. Our method achieves robustness to driving noise,
which is the typical problem of continuous dynamical decoupling schemes.
 Three level-systems are widely available and appear
in many atomic and solid state systems, such as trapped ions, rare-earth ions,
defect centers, and in particular, the NV-center in diamond.
This scheme is applicable to both optical and microwave configurations.
The fact that only the three-level system is manipulated
facilitates the realization of the fully robust qubit and its integration
in the target application. Moreover, the construction by off-resonant driving fields
enables the implementation of fast, simple qubit gates. Our scheme is therefore aimed at enhancing the performance of
a wide range of tasks in the fields of quantum information science and quantum
technologies, and in particular, quantum sensing, where due to the off-resonance construction, our scheme constitutes a
novel method for sensing high frequency signals. 	

\begin{figure}[t]
\centering{}\includegraphics[width=0.50\textwidth]{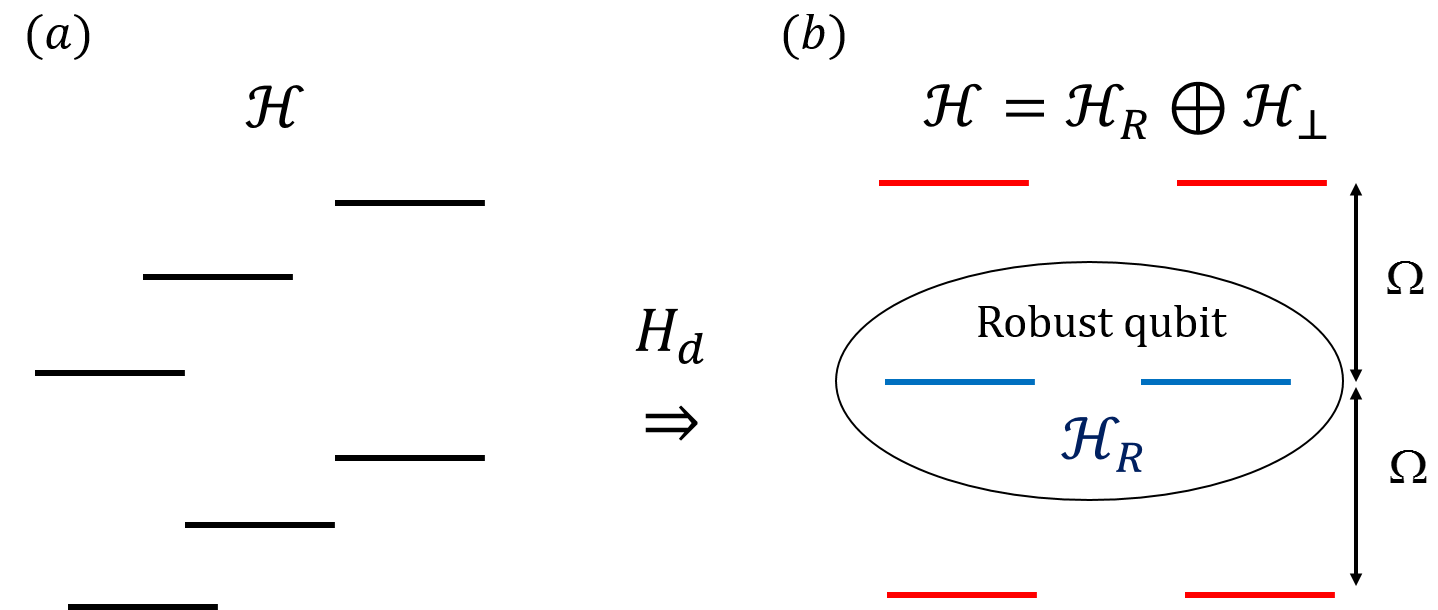}\protect\caption{\textbf{Fully robust qubit}. By the application of continuous driving fields we create a robust qubit subspace. Magnetic noise and power fluctuations of the driving fields do not operate within the robust qubit subspace. (a) Bare states, $H_d$  (driving Hamiltonian). (b) Fully robust qubit (blue), $\Omega$ (smallest energy gap between the robust qubit states and  non-robust states.}
\label{FR-0}
\end{figure}

\section{Fully robust qubit}

We start with an explicit definition of a fully robust qubit  \cite{ADR}. Let us  denote by $\left\{ \left|R_{i}\right\rangle \right\} $ the robust qubit states.
In what follows $H_{d}$ is the (continuous) driving Hamiltonian, $\mathcal{H}_{R}$ is the Hilbert subspace of the fully robust qubit,
and $\mathcal{H}_{\perp}$ is the complementary Hilbert space, that is, $\mathcal{H}=\mathcal{H}_{R}\oplus\mathcal{H}_{\perp}$.
We define the fully robust qubit  by (See Fig. \ref{FR-0})

\begin{eqnarray}
\left\langle R_{i}\right|S_{z}\left|R_{j}\right\rangle =0 &  & \qquad\forall i,j, \label{FRD:1} \\
H_d \ket{R_i} = \lambda^{R} \ket{R_i} &  & \qquad\forall i.\label{FRD:2}
\end{eqnarray}
The first equation ensures that magnetic noise does not operate within
the subspace of the fully robust qubit; the noise can only cause transitions between
a robust state and a state in the complementary
subspace. We assume (by construction) that the energy of all states in $\mathcal{H}_{R}$ is far from the energy
of the states in $\mathcal{H}_{\perp}$. More specifically, we assume that $\nu=\min_{i} |\lambda_{i}^{\perp}-\lambda^{R}|$, where
$\lambda^{R}$ ($\lambda_{i}^{\perp}$) is an eigenvalue of an eigenstate in $\mathcal{H}_{R}$ ($\mathcal{H}_{\perp}$), is much larger
than the characteristic frequency of the noise, as in this case the
lifetime $T_{1}$ would be inversely proportional to the power spectrum of the noise  at  $\nu$. This ensures that
the rate of transitions from $\mathcal{H}_{R}$ to $\mathcal{H}_{\perp}$ due to magnetic noise is
negligible.

The second equation indicates that the robust states do not collect
a relative dynamical phase due to $H_{d}$, and are therefore  immune to noise
originating from $H_{d}$. Power fluctuations of the driving fields result in  identical
energy fluctuations of the robust states.

To summarize, the first equation ensures that the robust states
are immune to external noise, while the second equation ensures that
the robust states are also immune to controller noise.

\begin{figure}[t]
\begin{centering}
\includegraphics[scale=0.36]{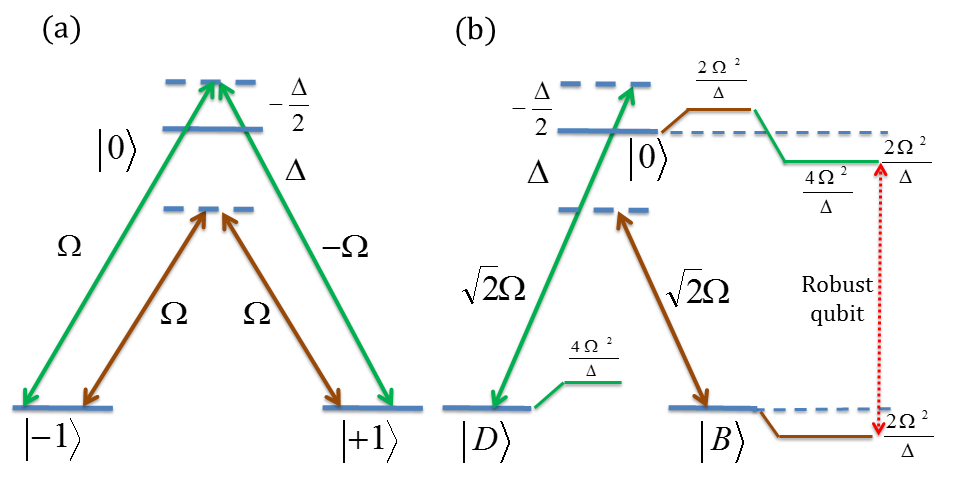}
\par\end{centering}

\protect\caption{\textbf{Fully robust qubit in a three-level system.} (a) Two $\Lambda$ systems are created
via the same level with two unequal detunings of opposite signs. (b)
The driving fields of the two $\Lambda$ systems result in Stark shifts
of the three levels, here described in the $\left\{ \ket{B},\ket{D},\ket{0}\right\}$
basis. In the case where the ratio between the red detuning and the blue
detuning is equal to 2 (and for the specific values of the Rabi frequencies),
the Stark shifts of the $\ket{B}$ and $\ket0$ states are identical. At the same time,
a large energy gap is formed between the $\ket{B}$ and $\ket{D}$ states.}

\label{Fig1}
\end{figure}

\section{Fully robust qubit in a three-level system} The rationale for the method is
illustrated in Fig. \ref{Fig1}. Driving a three-level system in a
$\Lambda$ configuration with large detunings results in Stark shifts
of all three levels. We design the driving fields; i.e., their Rabi
frequencies and detunings, in such a way that the new eigenstates are decoupled, in first order, from
the external magnetic field (see Eq. \ref{FRD:1}). In addition,
up to the second order, two of the eigenstates have an identical Stark shift (see Eq. \ref{FRD:2});
hence, fluctuations in the energy gap between them are mitigated since
noise in the driving fields will cause only fluctuations due to the higher order terms of the Stark shifts.
Specifically, we consider driving fields in two $\Lambda$ configurations. In one $\Lambda$ configuration the driving fields are red detuned and in the second $\Lambda$ configuration the driving fields are blue detuned. Denote by $\Omega$ the Rabi frequency of the driving fields, and by $\Delta$ the detuning. The red detuned driving fields, which correspond to (in the interaction picture (IP))
\begin{equation}\label{Hred}
  H^{red}=\Omega\left(\ket{0}\bra{-1}e^{-i\Delta t}+\ket{0}\bra{+1}e^{-i\Delta t}\right)+h.c.,
\end{equation}
result in the effective Hamiltonian \cite{James}
\begin{equation}\label{Hredeff}
  H_{eff}^{red}=-\frac{\Omega^{2}}{\Delta}\left(2S_{x}^{2}+4S_{z}^{2}-4\mathbf{1}\right).
\end{equation}
Similarly, the blue detuned driving fields, which correspond to
\begin{equation}\label{Hblue}
  H^{blue}=\Omega\left(\ket{0}\bra{-1}e^{+i\frac{\Delta}{2}t}-\ket{0}\bra{+1}e^{+i\frac{\Delta}{2}t}\right)+h.c.,
\end{equation}
result in the effective Hamiltonian
\begin{equation}\label{Hblueeff}
  H_{eff}^{blue}=-\frac{\Omega^{2}}{\Delta}\left(4S_{x}^{2}-4S_{z}^{2}\right).
\end{equation}
Our construction therefore results in the effective Hamiltonian
\begin{equation}\label{Heff}
  H_{eff}=H_{eff}^{red}+H_{eff}^{blue}=-\frac{\Omega^{2}}{\Delta}\left(6S_{x}^{2}-4\mathbf{1}\right),
\end{equation}
whose $\ket{B}=\frac{1}{\sqrt{2}}\left(\ket{+1}+\ket{-1}\right)$ and $\ket{0}$ eigenstates have a zero first order
Zeeman shift and identical energies. Hence, the two requirements for a fully robust qubit, Eq. \ref{FRD:1} and Eq. \ref{FRD:2},
are fulfilled by the $\ket{B}$ and $\ket{0}$ states (with  $H_d=H_{eff}$). Viewed in the $\left\{ \ket{B},\ket{D},\ket{0}\right\} $
basis, where $\ket{D}=\frac{1}{\sqrt{2}}\left(\ket{+1}-\ket{-1}\right)$, the red detuned driving fields induce a positive (negative)
Stark shift to the $\ket{0}$ ($\ket{B}$) state, while the blue detuned
driving fields induce a negative (positive) Stark shift to the $\ket{0}$
($\ket{D}$) state. The driving fields are therefore tuned such that
the total Stark shift of the $\ket{0}$ state will be equal to the
Stark shift of the $\ket{B}$ state (see Fig. \ref{Fig1}).

We assume a zero-field splitting between the $\ket{0}$ and $\ket{\pm1}$ states.  In case that the  $\ket{\pm1}$ states are split, due to a static magnetic field, the on-resonance frequencies of the $\ket{-1}\leftrightarrow\ket{0}$ and $\ket{+1}\leftrightarrow\ket{0}$
transitions are not identical, therefore, we consider the regime
where $g\mu_{B}B\gg\Delta\gg\Omega$. Hence, each $\Lambda$
system requires two different (phase-matched) driving fields and, in the case of a microwave implementation (with linear polarizations),
corrections on the order of $\sim\frac{\Omega^{2}}{g\mu_{B}B}$ are
introduced.

\section{Robustness} We first analyze the robustness of the scheme to environmental
and controller noise, which are extremely crucial to the NV-based implementation,
and then refer to possible errors in the general experimental set-up.

With respect to environmental noise, dephasing of the
dressed states is caused by two factors. The first source of
dephasing is the high order coupling to the external magnetic
field. By construction, the first order coupling is eliminated, but higher order terms remain. This can be grasped by moving to the time independent frame of the dressed states.
In the lab frame, and in the basis of the bare states, the Hamiltonian of the noise is given by
\begin{equation}\label{Hnoise}
  H_{noise}=g\mu_{B}B(t)S_{z},
\end{equation}
where $B(t)$ is a randomly fluctuating magnetic field. Moving to the IP
with respect to the energies of the bare states, and then moving to the basis of the dressed states,
$H_{noise}$ is transformed to
\begin{equation}\label{Hnoise1}
  H_{noise}^{I}=g\mu_{B}B(t)(\ketbradif{B}{D}+h.c.).
\end{equation}
We continue by moving to the time independent frame; that is, to the IP with respect to
$H_{0}^{I}= -\Delta \ketbra{B} + \frac{\Delta}{2}\ketbra{D}$.  This results in
\begin{equation}\label{Hnoise2}
  H_{noise}^{II}=g\mu_{B}B(t)(\ketbradif{B}{D} e^{-i\frac{3}{2}\Delta t}+h.c.).
\end{equation}
The Stark shifts obtained by the driving fields are accompanied by a small amplitude mixing
between the ideal $\{\ket{0}, \ket{B}, \ket{D}\}$ states (i.e., the exact eigenstates), which means
that $H_{noise}^{II}$ is further (slightly) rotated to have both diagonal and other off-diagonal terms.
However, due to the high detuning of $\sim \frac{3}{2}\Delta$, the effect of all of these contributions is negligible.
Therefore, the significant effect of the noise is due to the coupling between the $\ket{B}$ and $\ket{D}$
states. In the first order, the noise induces a longitudinal relaxation (decay) rate of $\sim S_{BB}(E_{BD})$, where  $S_{BB}$ is the power spectrum of the noise, and $E_{BD}$ is  the energy gap between the  $\ket{B}$ and $\ket{D}$ states. Hence, a large $E_{BD}$ ensures that the longitudinal relaxation rate is negligible ($\sim S_{BB}(E_{BD}) \ll \frac{1}{T_{1}}$). In this case, the noise does not induce transitions between the $\ket{B}$ and $\ket{D}$ states, but does result in a second order fluctuating phase shift of $\sim \frac{(g\mu_{B}B(t))^2}{E_{BD}}$. The resulting dephasing rate is considerably diminished with an increasing $E_{BD}$  (see Appendix A).
The second source of dephasing is due to the counter-rotating terms of the driving fields, which induce
minor mixing between the $\ket{B}$ and $\ket{D}$ states via a Raman transition. In case that the $\ket{\pm 1}$ states are Zeeman sub-levels, this results in an additional mixing term of $\sim(\frac{\Omega^{2}}{g\mu_{B}B}) S_{z}$ in the effective Hamiltonian of Eq. (\ref{Heff}),
and the mixing is of the order of $\sim(\frac{\Omega^{2}}{g\mu_{B}B})/(\frac{\Omega^{2}}{\Delta})=\frac{\Delta}{g\mu_{B}B}$. This implies a dephasing rate of $\sim \frac{\Delta}{g\mu_{B}B} S_{BB}(0)$, which is greatly suppressed by enlarging the Zeeman splitting.

Regarding controller noise, in an ideal construction the (second order) Stark shifts of the $\ket{B}$ and $\ket{0}$ states are identical and therefore immunity  to controller noise is obtained. However, while we can fix the second order Stark shifts to be identical, the fourth order terms might not be negligible, and in this case will introduce an energy gap between the $\ket{B}$ and $\ket{0}$ states. Fluctuations of this energy gap, due to driving amplitude noise,  can be significantly reduced by either an exact calculation of the fourth order energy shifts, or a numerical search for the point of a non-zero second order shift, which is robust to driving fluctuations \cite{DrivingNoise}.
In Appendix B we show how robustness could be further improved by utilizing a double-drive, where the first drive is on-resonance and the second drive is off-resonance.

For the case of an NV-center in diamond, which we analyzed in detail (see below),
our scheme achieves a significant improvement in the coherence time under
realistic conditions that take into account both environmental noise and power fluctuations of driving fields.

The robustness of the scheme may also be affected by errors in the experimental set-up.
An uncertainty, or a drift, of the static magnetic field, $\delta B_{z}$, shifts
the bare $\ket{+1}$ and $\ket{-1}$ states, and therefore introduces
two-photon detunings. Compared to the effect of the fluctuating magnetic noise, the dominant effect here is a first order effect. The coupling between the $\ket{B}$ and $\ket{D}$ states results in an amplitude mixing and the $\ket{B}$ state is modified to $\ket{\tilde{B}}\sim\ket{B}+\frac{g\mu_{B}\delta B_{z}\Delta}{\Omega^{2}}\ket{D}$. Hence,   $\delta B_{z}$ inflicts a dephasing rate of $\sim\frac{g\mu_{B}\delta B_{z}\Delta}{\Omega^{2}} S_{BB}(0)$.
This dephasing rate, however, remains negligible as long as the energy
gap between the dressed states is much larger than the magnetic field
uncertainty; that is, $\frac{\Omega^{2}}{\Delta}\gg g\mu_{B}\delta B_{z}$.
In addition, there can be relative amplitude and relative phase errors
between the two driving fields of a $\Lambda$ system. In both cases, a relative
error of $\epsilon$ will introduce an amplitude mixing of $\sim \frac{\epsilon \Omega}{\Delta}$ and an energy shift of $\sim  \frac{\epsilon^{2} \Omega^{2}}{\Delta}$. For example, a relative amplitude error of $\epsilon$ in the red detuned $\Lambda$ system introduces (in the IP) the coupling term
$\epsilon \Omega(\ketbradif{0}{D} e^{-i\frac{3}{2}\Delta t}+h.c.)$, which results in an amplitude mixing of $\sim \frac{\epsilon \Omega}{\Delta}$ between the $\ket{0}$ and $\ket{D}$ states. Since the magnetic noise rotates at the same frequency as this coupling term (see Eq. \ref{Hnoise2}) and because there is an amplitude mixing of $\sim \frac{\Omega}{\Delta}$ between the $\ket{0}$ and $\ket{B}$ states, we have that $\bra{0}S_{z}\ket{0}\sim (\frac{\Omega}{\Delta})^2 \epsilon$, and hence, the inflicted dephasing rate due to a relative amplitude error of $\epsilon$ is $\sim(\frac{\Omega}{\Delta})^2 \epsilon S_{BB}(0).$

\begin{figure}[t]
\begin{centering}
\includegraphics[scale=0.36]{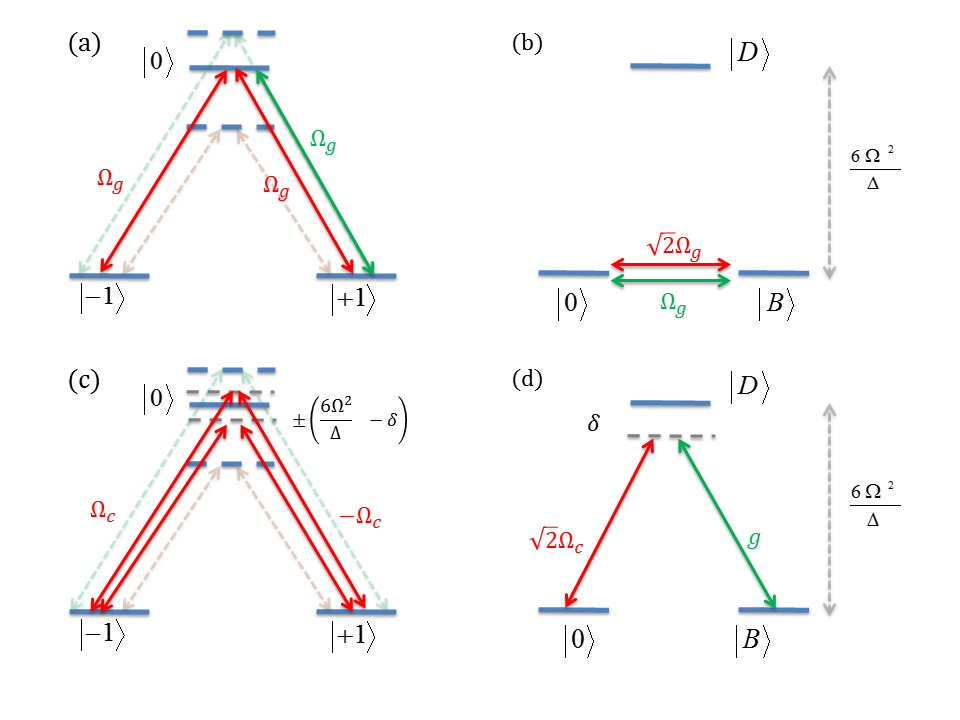}
\par\end{centering}

\protect\caption{\textbf{Single qubit gate and sensing.} A single qubit gate in the
bare states basis (a), and in the dressed states basis (b).
Red (green) arrows correspond to a gate with $\Omega_{g}\ll\Delta$ ($\Omega_{g}\ll\frac{\Omega^{2}}{\Delta}$).
The green gate enables the sensing of high frequency fields.
(c) Control field used for the sensing of low frequency fields via a Raman transition in the bare
states basis. (d) The sensing Raman transition in the dressed states
basis, where $g$ denotes the sensing field. Dashed
arrows in (a) and (c) represent the dressing driving fields.}

\label{Fig2}
\end{figure}

\section{Single qubit gates} In this section we show how protected qubit
gates can be implemented and discuss their application for sensing. A $\sigma_{x}$ gate can be realized
by driving the $\ket{B}\leftrightarrow\ket{0}$ transition on resonance with
\begin{equation}\label{Gate}
  H_{x}=\Omega_{g}\left(\cos\left(\omega_{-1,0}t\right)\ket{0}\bra{-1}+\cos\left(\omega_{+1,0}t\right)\ket{0}\bra{+1}\right)+h.c..
\end{equation}
Note that while a concatenated  on-resonance driving scheme allows for slow
gates with $\Omega_{g}\ll\Omega_{n}$, where $\Omega_{n}$ is the Rabi frequency of the last driving field, our
method enables fast gates, where $\Omega_{g}$ is limited solely by the detuning, $\Omega_{g}\ll\Delta$ (see Fig.
$\ref{Fig2}$).
A $\sigma_{y}$ gate can be realized by introducing
a phase shift of $\frac{\pi}{2}$ in the driving frequency with respect
to the driving frequency of the $\sigma_{x}$ gate. Alternatively,
one can start with a polarization that corresponds to the $\sigma_{y}$
gate, and then add the $\frac{\pi}{2}$ phase shift to get
the $\sigma_{x}$ gate. These realizations of $\sigma_{x}$ or $\sigma_{y}$
gates require two (phase-matched) driving fields, which only couple
the $\ket{B}$ state to the $\ket{0}$ state (similar to the dressing fields). A simpler implementation
of the gates can be achieved by employing only one of the driving fields.
However, as this driving field couples both the $\ket{B}$ and $\ket{D}$
states to the $\ket{0}$ state, $\Omega_{g}$ is limited by $\Omega_{g}\ll\frac{\Omega^{2}}{\Delta}$.

\section{Sensing} Sensing of high frequency signals is of great importance, especially in the case of classical fields sensing \cite{Chipaux,Kolkowitz}, in detection of electron spins in solids \cite{Hall2016} and NMR \cite{Kimmich}.
To the best of our knowledge, to date, dynamical decoupling techniques have not been incorporated
in sensing schemes of high frequency signals,
which are therefore limited by  $T_{2}^{*}$.
 Our scheme enables enhanced sensing of high frequency
AC signals, where a signal induces rotations of the fully robust qubit.
This can be accomplished by tuning the frequency of the $\ket{B}\leftrightarrow\ket{0}$
transition to the sensing field frequency, as in this
case the frequency corresponds to the energy gap between the bare $\ket{0}$
and $\ket{\pm1}$ states. Since the sensing sensitivity scales, in the shot noise limit, like $\sqrt{T_{2}}$,
for the case of sensing with an NV-center our scheme predicts an improvement of
$\sim1$ order of magnitude in sensitivity.

Sensing of AC signals with lower frequencies can by done by a Raman
transition. We assume that the AC signal corresponds to a $\sigma_{z}$
operation, which couples the $\ket{B}$ and $\ket{D}$ states,
and its amplitude is denoted by $g$. A Raman transition between the
$\ket{B}$ and $\ket{0}$ states is achieved by adding a control field
whose frequency is tuned to match the same detuning as that of the
AC signal, so a one-photon detuning is obtained (See Fig. $\ref{Fig2}$
(c) ,(d)). Full oscillation will then be observed whenever $\sqrt{2}\Omega_{c}=g$,
where $\Omega_{c}$ is the Rabi frequency of the control field. In
this case the sensing sensitivity is limited by the fluctuations of
the (dressing) Rabi frequency, $\Omega$, which results in fluctuations
of the one-photon detuning, $\delta$. Ideally, the sensitivity scales
like $\frac{\delta}{g}\sqrt{T_{2}^{\Omega}}$, where $T_{2}^{\Omega}$
is the coherence time induced by the Rabi frequency fluctuations.
Note that the sensitivity of  low-frequency signal sensing using
the bare state scales like $\sqrt{T_{2}^{*}}$, while a scaling of $\sqrt{T_{2}^{\Omega}}$
is obtained by utilizing the $\ket{B}\leftrightarrow\ket{D}$
transition of the dressed states.

\begin{figure}[t,l]
\begin{centering}
\includegraphics[scale=0.36]{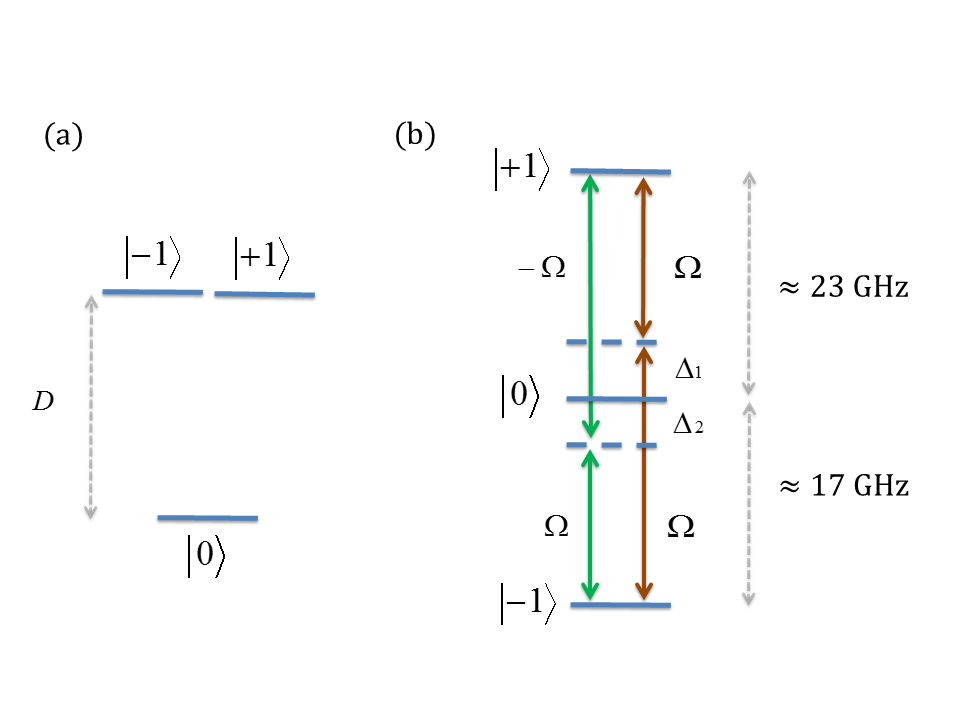}
\par\end{centering}

\protect\caption{\textbf{Implementation with the NV-center.} Ground state of the NV-center.
(a) Without a static magnetic field. (b) With a static magnetic field and driving fields.
The ratio of $\Delta_{1}$ to $\Delta_{2}$ is chosen such that robustness to power fluctuations of driving fields is achieved.}

\label{Fig3}
\end{figure}

\section{Implementation with the NV-center in diamond} The electronic ground state of
the NV-center is a spin $1$ state, where the $\ket{\pm 1}$ states are separated from
the $\ket{0}$ state by a zero-field splitting of $D=2.87$ GHz \cite{Fedor1,Fedor2} (see Fig. \ref{Fig3} (a)). We
consider a static magnetic field, which is applied along the NV axis, such that $g\mu_{B}B\approx20$ GHz
(note that a larger Zeeman spitting would result in a better decoupling from the magnetic noise).
In this case level-crossing occurs and the energy gaps between  $\ket{0}$ and  $\ket{\pm 1}$ correspond
to $\omega_{0,+1}\approx23$ GHz and  $\omega_{0,-1}\approx17$ GHz (see Fig. \ref{Fig3} (b)). We assume
that to have a good decoupling of the robust qubit and the magnetic noise we need to create
an energy gap of $\gtrsim 10$ MHz between the $\ket{B}$ and $\ket{D}$ states; hence,
we set $\Omega=70$ MHz \cite{Fuchs, Shin}. This implies that the conditions for an ideal construction, $g\mu_{B}B,\omega_{0}\gg\Delta\gg\Omega$, are not fully satisfied and therefore the Stark shifts will have contributions from all driving fields as
well as from the counter-rotating terms. The Hamiltonian of the system is given by
\begin{eqnarray}
  H &=& \omega_{0} S_{z}^{2}+\omega_{B} S_{z}\nonumber \\
    &+& \Omega S_{x}\biggl(\cos\left[\left(\omega_{0}+\omega_{B}-\Delta_{1}\right)t\right]+\cos\left[\left(\omega_{0}-\omega_{B}-\Delta_{1}\right)t\right]\nonumber \\
    &+&
    \cos\left[\left(\omega_{0}+\omega_{B}+\Delta_{2}\right)t\right]-\cos\left[\left(\omega_{0}-\omega_{B}+\Delta_{2}\right)t\right]\biggr),
\end{eqnarray}
where $\omega_{0}=D$ and $\omega_{B}=g\mu_{B}B$. By moving to the
IP with respect to $H_{0}=\omega_{0} S_{z}^{2}+\omega_{B} S_{z}$
(but not taking the rotating-wave approximation (RWA)), and then moving to the ideal dressed states
basis , $\{\ket{0},\ket{B},\ket{D}\}$ we obtain $H_{I}=Ue^{iH_{0}t}He^{-iH_{0}t}U^{\dagger}$,
from which we calculate, in the ideal dressed states basis, the energy shifts of the dressed states (up
to the second order) \cite{James,James1}. The energy shifts are given by  \cite{James2}
\begin{eqnarray}
\Delta E_{B} &=& \frac{1}{8}\Omega^{2}\left(\frac{4}{\text{\ensuremath{\Delta_{1}}}}+\frac{4}{2\text{\ensuremath{\omega_{0}}}-\text{\ensuremath{\Delta_{1}}}}\right.\nonumber\\
  &+& \frac{1}{2\omega_{B}+\text{\ensuremath{\Delta_{1}}}}-\frac{1}{2\omega_{B}-\text{\ensuremath{\Delta_{1}}}}+\frac{1}{2\omega_{B}-\text{\ensuremath{\Delta_{2}}}}-\frac{1}{2\omega_{B}+\text{\ensuremath{\Delta_{2}}}} \nonumber\\
  &+& \frac{1}{-2\omega_{B}-\text{\ensuremath{\Delta_{1}}}+2\text{\ensuremath{\omega_{0}}}}+\frac{1}{2\omega_{B}-\text{\ensuremath{\Delta_{1}}}+2\text{\ensuremath{\omega_{0}}}}\nonumber\\
  &+& \left.\frac{1}{-2\omega_{B}+\text{\ensuremath{\Delta_{2}}}+2\omega_{0}}+\frac{1}{2\omega_{B}+\Delta_{2}+2\text{\ensuremath{\omega_{0}}}}\right),\\
\Delta E_{D} &=& \frac{1}{8}\Omega^{2}\left(-\frac{4}{\text{\text{\ensuremath{\Delta_{2}}}}}+\frac{4}{2\omega_{0}+\text{\ensuremath{\Delta_{2}}}}\right.\nonumber\\
&+& \frac{1}{2\omega_{B}-\text{\text{\ensuremath{\Delta_{2}}}}}-\frac{1}{2\omega_{B}+\text{\ensuremath{\Delta_{2}}}}+\frac{1}{2\omega_{B}+\text{\text{\ensuremath{\Delta_{1}}}}}-\frac{1}{2\omega_{B}-\text{\text{\ensuremath{\Delta_{1}}}}}\nonumber\\
&+& \frac{1}{-2\omega_{B}+\Delta_{2}+2\text{\ensuremath{\omega_{0}}}}+\frac{1}{2\omega_{B}+\Delta_{2}+2\text{\ensuremath{\omega_{0}}}}\nonumber\\
&+& \left.\frac{1}{-2\omega_{B}-\text{\ensuremath{\Delta_{1}}}+2\text{\ensuremath{\omega_{0}}}}+\frac{1}{2\omega_{B}-\text{\ensuremath{\Delta_{1}}}+2\omega_{0}}\right),\\
\Delta E_{0} & = & -\Delta E_{B}-\Delta E_{D}.
\end{eqnarray}
In an ideal scenario
the terms $\sim\frac{\Omega^{2}}{\text{\ensuremath{\omega_{0}}}},\frac{\Omega^{2}}{\text{\ensuremath{\omega_{B}}}}$
would be negligible, and hence, the requirement $\Delta E_{0}=\Delta E_{B}$ would imply  $\Delta_{2}=\frac{\Delta_{1}}{2}$.

In order to achieve an energy gap of $\gtrsim 10$ MHz between the $\ket{B}$ and $\ket{D}$ states, together with $\Omega=70$ MHz,
we also set $\Delta_{1}=500$ MHz. For $\Delta E_{0}=\Delta E_{B}$, the energy gap between the $\ket{0}$ and $\ket{B}$ states,
due to the fourth order energy shifts, is $E_{0B}=E_{0} - E_{B}\approx 0.25$ MHz, which means that driving fluctuations will impose a limitation on the coherence time. We therefore tune the energy shifts to a robust point at which $\Delta E_{0}-\Delta E_{B}\approx 0.63$ MHz, and $ E_{0B}\approx0.315$ MHz \cite{DrivingNoise}. In this case we have that  $\Delta_{1}=500$ MHz,
$\Delta_{2}\approx209$ MHz, and $E_{BD}\approx17.96$ MHz.

\begin{figure}[t,l]
\includegraphics[width=0.51\textwidth]{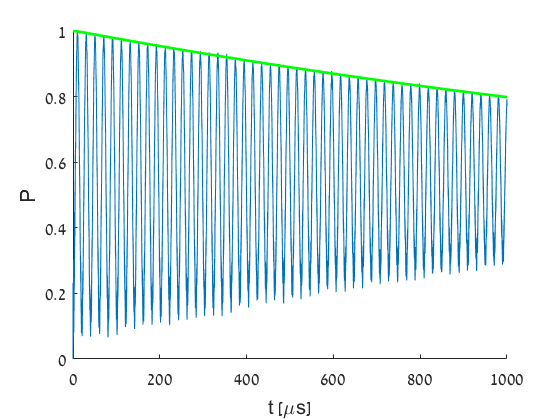}

\protect\caption{\textbf{Coherence time.}
Simulation of an NV-center implementation of a fully robust qubit
where $T_{2}^{*}=5$ $\mu$s, $\Omega=70$ MHz, $\Delta_{1}=500$ MHz and $\Delta_{2}=209$ MHz.
The graph is a result of average over $200$ trails, and shows oscillations between the $\ket{\psi_{\pm}}=\frac{1}{\sqrt{2}}\left(\ket{0}\pm\ket{B}\right)$ states.  The theoretical dephasing rate is plotted in green and corresponds to a coherence time of $T_{2}\simeq 1820\:\mu s$.}

\label{Fig4}
\end{figure}

We verified the robustness of this scheme by simulating its performance when the NV spin was subject to magnetic noise and driving fluctuations. We modelled the magnetic noise, $B\left(t\right)$, as an Ornstein-Uhlenbeck process \cite{OU1,OU2} with a zero expectation value, $\left\langle B\left(t\right)\right\rangle =0$,
and a correlation function $\left\langle B\left(t\right)B\left(t^{'}\right)\right\rangle =\frac{c\tau}{2}e^{-\gamma\left|t-t^{'}\right|}$.
An exact simulation algorithm \cite{OU3} was employed to realize the
Ornstein-Uhlenbeck process, which according to
\begin{equation}
B(t+\Delta t)=B(t)e^{-\frac{\Delta t}{\tau}}+n\sqrt{\frac{c\tau}{2}\left(1-e^{-\frac{2\Delta t}{\tau}}\right)},
\end{equation}
where $n$ is a unit Gaussian random number. We took the pure dephasing
time to be $T_{2}^{*}=5\:\mu s,$ and the correlation time of the noise
was set to $\tau=\frac{1}{\gamma}=15\:\mu s$ \cite{Lange1,BarGil1}, where the diffusion constant is
given by $c\approx\frac{4}{T_{2}^{*}{}^{2}\tau}.$  Driving fluctuations were also modelled by an an Ornstein-Uhlenbeck process  with a zero expectation value. We chose a correlation time of
$\tau_{\Omega}=500\:\mu s$, and a relative amplitude error of $\delta_{\Omega}=0.5\%$  so the diffusion constant is given by $c_{\Omega}=2\delta_{\Omega} \big/ \tau_{\Omega}$.
Fig. \ref{Fig4} presents the outcome of the simulation of the fully robust qubit under the effect of magnetic and driving noise.
The plot shows oscillations between the $\ket{\psi_{\pm}}=\frac{1}{\sqrt{2}}\left(\ket{0}\pm\ket{B}\right)$ states
averaged over $200$ trials. The oscillations are not symmetric because  fast oscillations due to counter-rotating terms at (local) minimum values of $P$ are averaged to zero. The simulation confirmed our estimation of $T_{2}\simeq 1820\:\mu s$, an improvement of more than 2 orders of magnitude in the coherence time, pushing it towards the lifetime limit. Note that the simulation does not take decoherence due to longitudinal spin relaxation (of the bare states) into account, which is given by $\Gamma_{2}=\frac{\Gamma_{1}}{2}$, where $\Gamma_{1}=\frac{1}{T_{1}}$ (since $S_{BB}(E_{BD}) \ll \Gamma_{1}$, the effect of the noise on the life time of the dressed states is negligible).

 The probability of remaining in the initial  $\ket{\psi_{+}}$ state is given by  (green line in Fig. \ref{Fig4})

\begin{equation}
P=\frac{1+|F(t)G(t)|e^{-\gamma_{m}t}e^{-(\gamma_{d}t)^2}}{2}, \label{Ptotal}
\end{equation}

where
\begin{eqnarray}
F\left(t,\Omega\right) & =\frac{\exp\left(\frac{\gamma t}{2}\right)}{\sqrt{\cosh\left(\frac{\xi t}{2}\right)+\frac{2\gamma}{\xi}\sinh\left(\frac{\xi t}{2}\right)}},\\
G\left(t,\Omega\right) & =\exp\left(\frac{2ig^{2}}{\Omega\left(2\gamma+\xi\coth\left(\frac{\xi t}{2}\right)\right)}\right),
\end{eqnarray}
and
\begin{equation}
\xi=\sqrt{4\gamma^{2}-\frac{16i\gamma g^{2}}{\Omega}}.
\end{equation}
$|F(t)G(t)|$ corresponds to the (second order) dephasing due to the coupling between $\ket{B}$ and $\ket{D}$ (see Appendix A),
$\gamma_{m}=\bra{B}S_{z}\ket{B}S_{BB}(0)$ is the (first order) dephasing rate due to the amplitude mixing between $\ket{B}$ and $\ket{D}$,
and $\gamma_{d}=\frac{\delta_{r}\delta_{\Omega}\Omega}{\sqrt{2}}$ is the dephasing rate due to driving fluctuations, where $\delta_{r}=\frac{|E_{0B}(\Omega+\delta \Omega)-E_{0B}(\Omega)|}{E_{0B}(\Omega)}$ . In our case we estimated that $\gamma_{m}\approx200$Hz, $\gamma_{d}=182$Hz, and the coherence time due to the coupling between $\ket{B}$ and $\ket{D}$ alone is $\simeq 3440\:\mu s$. In Appendix C we show the effect of these different sources of noise on the coherence time, which together result in $T_{2} = 1820\:\mu$s.

\begin{figure}[t,r]
\includegraphics[width=0.48\textwidth]{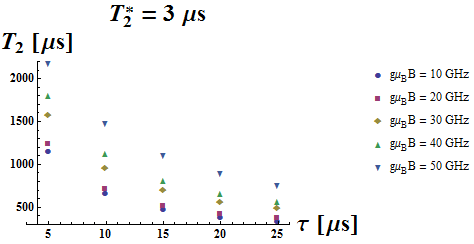}

\protect\caption{\textbf{Lower bound estimation of $T_{2}$.}
A theoretical (non-optimized) estimation of the coherence times for the case of
$T_{2}^{*}=3$ $\mu$s as function of the correlation time of the noise and for various values of the Zeeman splitting.}

\label{Fig5}
\end{figure}

We used our theoretical model to estimate the achievable coherence times in different scenarios. Fig. \ref{Fig5} shows the estimated
coherence times for the case of $T_{2}^{*}=3\:\mu s$ as function of the correlation time of the noise and for various values of the Zeeman splitting.
The parameters chosen for these estimations (see Appendix D) were not optimized and thus the obtained $T_{2}$ times constitute a lower bound estimation. Nevertheless, the estimations imply that a significant improvement in the coherence time can be achieved under even more severe conditions.

\
\section{Conclusion} We presented a new method that enables the construction
of a fully robust qubit utilizing a three-level system alone. By the application
of off-resonance continuous driving fields in a $\Lambda$ configuration, robustness
to both external and controller noise is achieved. We analyzed the NV-center based
implementation of the scheme and showed that with current state-of-the-art experimental setups
the scheme enables an improvement of more than two orders of magnitude in the coherence time.
Moreover, since the scheme allows for fast gates, it is advantageous with respect to on-resonance driving schemes,
since more qubit operations in a given $T_{2}$ time interval can be performed.
Our analysis of the NV-center based implementation considered linearly polarized fields.
The performance of the scheme is likely to be further improved by the application of  circularly polarized fields \cite{London}.
This scheme is relevant to many tasks in the fields of quantum information science and quantum technologies,
and in particular to quantum sensing of high frequency signals.  The utilization of off-resonance driving fields makes the scheme more robust to an inhomogeneous broadening than schemes that use (continuous) on-resonance driving fields, and hence, it is more attractive for ensemble-based sensing.
Our scheme is expected to perform even better in the optical regime, where large energy gaps, stronger driving fields, and polarization dependent transitions  allow for much
smaller mixing amplitudes between the ideal dressed states.
Although here we considered  the case of a spin $1$ system, the scheme is also applicable to
systems of half-integer spins. For example, in the case of the calcium ion, $^{40}\textrm{Ca}^{+}$, one
could consider a $\Lambda$ system composed of the $\ket{S_{1/2};+1/2}$, $\ket{D_{3/2};-1/2}$, $\ket{D_{3/2};+3/2}$ states.
In this case a fully robust optical qubit can be realized with $\ket{0}=\ket{S_{1/2};+1/2}$ and $\ket{B}=\sqrt{\frac{1}{8}}\ket{D_{3/2};-1/2}+\sqrt{\frac{7}{8}}\ket{D_{3/2};+3/2}$.

\begin{acknowledgments}
A. R. acknowledges the support of the Israel Science Foundation (grant no. 039-8823), the support of the European commission (STReP EQUAM Grant Agreement No. 323714), EU Project DIADEMS, the Marie Curie Career Integration Grant (CIG) IonQuanSense(321798), the Niedersachsen-Israeli Research Cooperation Program and DIP program (FO 703/2-1). This work was [partially] supported by the US Army Research Office under Contract W911NF-15-1-0250. This project has received funding from the European Union as part of the Horizon 2020 research and innovation program under grant agreement No. 667192.  F. J. acknowledges the support of ERC, EU projects SIQS and DIADEMS, VW Stiftung, DFG and BMBF.
\end{acknowledgments}

\appendix

\section{}

Here we analyze the dephasing of a strongly driven system.
We consider the case of a two-level system (TLS) under a single on-resonance
driving and magnetic noise. The Hamiltonian is given by
\[
H=\frac{\omega_{0}}{2}\sigma_{z}+\Omega\cos\left(\omega_{0}t\right)\sigma_{x}+B\left(t\right)\sigma_{z},
\]
where $B(t)$ is the random magnetic noise (here in units of frequency). Moving to the interaction
picture (IP) with respect to $H_{0}=\frac{\omega_{0}}{2}\sigma_{z}$,
taking the rotating-wave-approximation (RWA), and moving to the dressed
states basis, we get that
\[
H_{I}=\frac{\Omega}{2}\sigma_{z}+B\left(t\right)\sigma_{x}.
\]

\begin{figure}[t]
\begin{centering}
\includegraphics[width=0.50\textwidth]{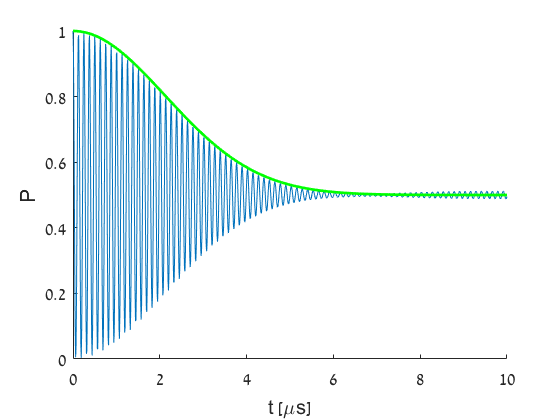}
\par\end{centering}
\protect\caption{Simulation of pure dephasing with no driving fields. $\frac{1+e^{-\frac{g^{2}t^{2}}{2}}}{2}$
is plotted in green $\left(g^{2}=\frac{2}{T_{2}^{*}{}^{2}}\right)$.}
\label{FigS1}
\end{figure}

In the regime of a strong driving field, $\Omega\gg\left|B\left(t\right)\right|$,
the time evolution of the dressed states can be simplified by the
adiabatic approximation and hence, the dressed states accumulate a
phase which is given by
\[
\phi\left(t\right)=\pm\frac{1}{2}\int_{0}^{t}dt^{'}\sqrt{4B^{2}\left(t^{'}\right)+\Omega^{2}}\approx\pm\frac{1}{2}\int_{0}^{t}dt^{'}\left(\Omega+\frac{2B^{2}\left(t^{'}\right)}{\Omega}\right).
\]
 We assume that $B\left(t\right)$ is an Ornstein-Uhlenbeck random
process \cite{OU1,OU2}, which is described by the stochastic differential equation
\[
dB_{t}=-\gamma B_{t}dt+c^{\frac{1}{2}}dW_{t},
\]
 where $\gamma=\frac{1}{\tau}$, $\tau$ and $c$ are the correlation
time and the diffusion constant of the noise, and $W_{t}$ is a Wiener
process. In this case $B^{2}\left(t\right)$ is known
as the square-root process, or the Cox-Ingersoll-Ross (CIR) process \cite{Carr},
whose stochastic differential equation is given by
\[
dB_{t}^{2}=\left(c-2\gamma B_{t}^{2}\right)dt+2c^{\frac{1}{2}}\sqrt{B_{t}^{2}}dW_{t}.
\]
 Denote the random phase by $\varphi\left(t\right)=\frac{1}{\Omega}\int_{0}^{t}dt^{'}B^{2}\left(t^{'}\right)$.
The characteristic function of the square-root process is explicitly
given by \cite{Carr, Dobrovitski}
\[
\left\langle e^{i\varphi\left(t\right)}\right\rangle _{B_{t}^{2}}=F\left(t,\Omega\right)G\left(t,\Omega\right),
\]
 where
\begin{align*}
F\left(t,\Omega\right) & =\frac{\exp\left(\frac{\gamma t}{2}\right)}{\sqrt{\cosh\left(\frac{\xi t}{2}\right)+\frac{2\gamma}{\xi}\sinh\left(\frac{\xi t}{2}\right)}},\\
G\left(t,\Omega\right) & =\exp\left(\frac{2ig^{2}}{\Omega\left(2\gamma+\xi\coth\left(\frac{\xi t}{2}\right)\right)}\right),
\end{align*}

\[
\xi=\sqrt{4\gamma^{2}-\frac{16i\gamma g^{2}}{\Omega}},
\]
 and we assume that $B^{2}(t=0)=\left\langle B^{2}(t)\right\rangle =g^{2}=\frac{c\tau}{2}$.

We therefore conclude that in the strong driving regime, the probability
to remain in the initial equal superposition state of the dressed
eigenstates is given by
\[
P_{\Omega}\left(t\right)=\frac{1+\left|F\left(t,\Omega\right)G\left(t,\Omega\right)\right|}{2}.
\]

\begin{figure}[t!]
 \includegraphics[width=0.50\textwidth]{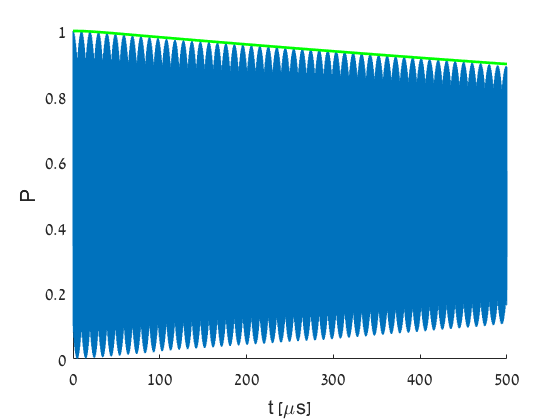}
\caption{Simulation of the coherence time under a driving of $\Omega=50\:\mathrm{MHz}$.
Average over 1000 trials. $P_{\Omega}\left(t\right)$ is plotted in
green.}
\label{FigS2}
\end{figure}

\begin{figure}[t!]
\includegraphics[width=0.50\textwidth]{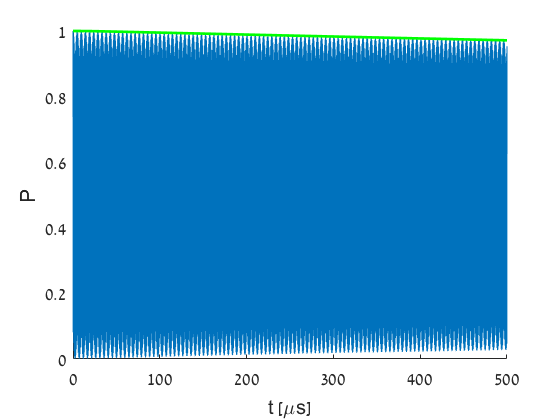}
\caption{Simulation of the coherence time under a driving of $\Omega=100\:\mathrm{MHz}$.
Average over 1000 trials. $P_{\Omega}\left(t\right)$ is plotted in
green.}
\label{FigS3}
\end{figure}

\begin{figure}[h]
\centering{}\includegraphics[width=0.50 \textwidth]{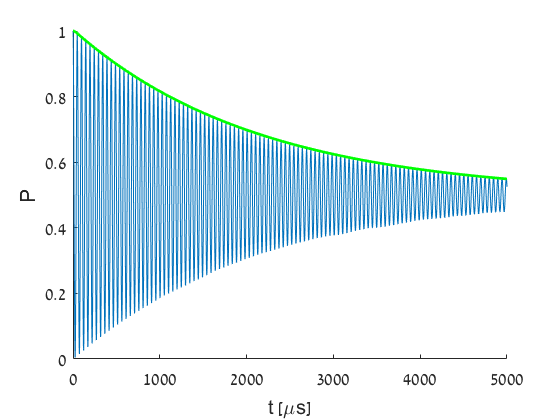}\protect\caption{Adiabatic approximation with  $\Omega=50\:\mathrm{MHz}$.
Numerical calculation of $P=\frac{1+\cos\left(\frac{1}{2}\int_{0}^{t}dt'\sqrt{4B^{2}\left(t'\right)+\Omega^{2}}\right)}{2}$.
Average over 10000 trials. $P_{\Omega}\left(t\right)$ is plotted in
green.}
\label{FigS4}
\end{figure}

\begin{figure}[t!]
\centering{}\includegraphics[width=0.50 \textwidth]{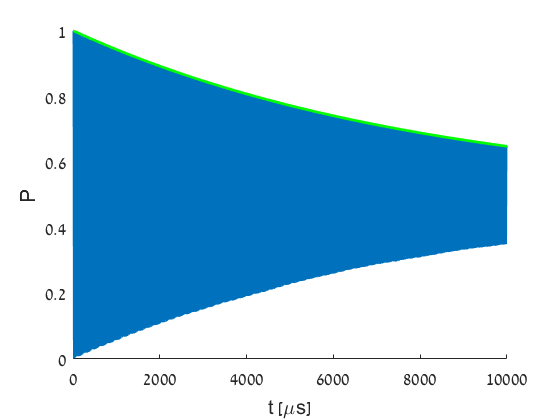}\protect\caption{Adiabatic approximation with $\Omega=100\:\mathrm{MHz}$.
Numerical calculation of $P=\frac{1+\cos\left(\frac{1}{2}\int_{0}^{t}dt'\sqrt{4B^{2}\left(t'\right)+\Omega^{2}}\right)}{2}$.
Average over 10000 trials. $P_{\Omega}\left(t\right)$ is plotted in
green.}
\label{FigS5}
\end{figure}

\begin{figure}[t]
\includegraphics[width=0.48\textwidth]{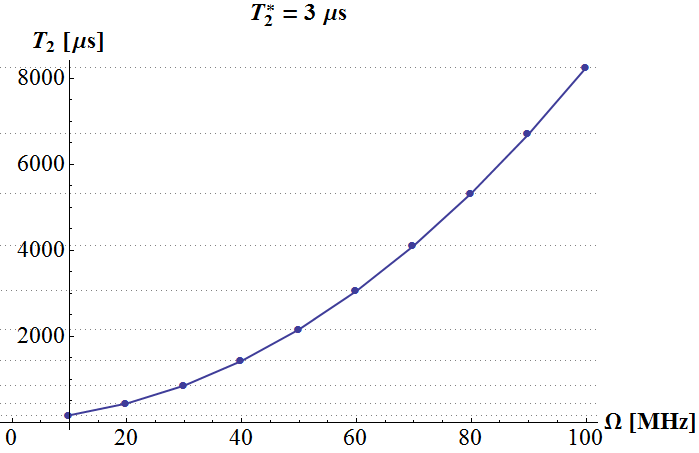}

\protect\caption{\textbf{$T_2$ as function of $\Omega$.}
The coherence times where deduced by setting $P_{\Omega}\left(t\right) = \frac{1+1/e}{2}$.}
\label{T2Omega}
\end{figure}

We numerically verified this by simulating the noise, $B\left(t\right)$,
as an Ornstein-Uhlenbeck process with a zero expectation value, $\left\langle B\left(t\right)\right\rangle =0$,
and a correlation function $\left\langle B\left(t\right)B\left(t^{'}\right)\right\rangle =\frac{c\tau}{2}e^{-\gamma\left|t-t^{'}\right|}$.
An exact simulation algorithm \cite{OU3} was employed to realize the
Ornstein-Uhlenbeck process, which according to
\[
B(t+\Delta t)=B(t)e^{-\frac{\Delta t}{\tau}}+n\sqrt{\frac{c\tau}{2}\left(1-e^{-\frac{2\Delta t}{\tau}}\right)},
\]
where $n$ is a unit Gaussian random number. We took the pure dephasing
time to be $T_{2}^{*}=3\:\mu s,$ and the correlation time of the noise
was set to $\tau=25\:\mu s$. The diffusion constant was therefore
given by $c\approx\frac{4}{T_{2}^{*}{}^{2}\tau}.$ In Fig. \ref{FigS1} the pure
dephasing (no driving) is plotted. Then, for two values of $\Omega$,
$\Omega=50$ MHz and $\Omega=100$ MHz we simulated the time evolution
of the TLS, which is initialized to $\left|\uparrow_{z}\right\rangle $
, the equal superposition of the dressed eigenstates. Fig. \ref{FigS2} and Fig. \ref{FigS3}
show the probability of remaining in the initial state as a function of
time. The analytical expression of $P_{\Omega}\left(t\right)$ is
plotted in green. In addition, we numerically calculated this probability,
which by the adiabatic approximation is given by $P=\frac{1+\cos\left(\frac{1}{2}\int_{0}^{t}dt'\sqrt{4B^{2}\left(t'\right)+\Omega^{2}}\right)}{2}$.
In Fig. \ref{FigS4} and Fig. \ref{FigS5} $P$ is plotted as a function of time and agrees with the analytical
expression of $P_{\Omega}\left(t\right)$, which is plotted in green. Increasing $\Omega$ increases $T_2$. Indeed, a $T_2=\{167,857,2163,4110,6707\}\: \mu$s is obtained with a driving of $\Omega=\{10,30,50,70,90\}$ MHz respectively. If Fig.\ref{T2Omega} we plot $T_2$ as function of $\Omega$.

\section{}

\begin{figure}[t!]
\begin{centering}
\includegraphics[scale=0.36]{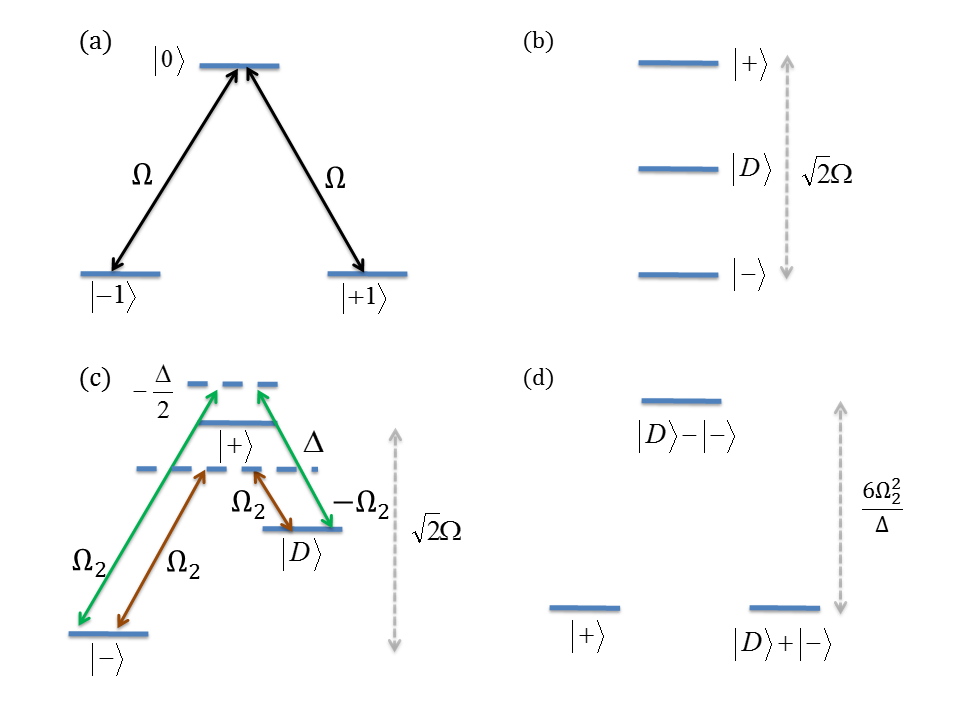}
\par\end{centering}

\protect\caption{\textbf{Improving robustness by a double-drive.} The first on-resonance
driving in the bare states basis (a), and the obtained dressed states
(b), which are immune to magnetic field fluctuations. (c) Off-resonance
driving fields in the dressed states basis and in the first IP, which
result in the effective $S_{x}^{2}$ Hamiltonian. (d) Doubly-dressed
states. The Stark shifts of the $\ket{+}$ and $\ket{D}+\ket{-}$
states are identical. }

\label{Fig6}
\end{figure}

The robustness
of the scheme to external noise depends on the energy gap between
the dressed states, and can, in principle, be improved by increasing
both the Rabi frequency of the driving fields and the detuning $\Delta$.
As these are limited, an improvement can be achieved by a double-drive,
where in the first drive on-resonance driving fields are applied.
The energy gap of the dressed states, which are immune to external
noise, is now $\sim\Omega$ (compared to an energy gap of $\frac{\Omega^{2}}{\Delta}$
in the case of a single off-resonance driving) (see Fig. $\ref{Fig5}$
(a),(b)). Next, we add off-resonance driving fields, which results in
an effective $S_{x}^{2}$ Hamiltonian of the dressed states, and thus
achieves robustness to controller noise as well (see Fig. $\ref{Fig6}$
(c),(d)). In the IP, and taking the RWA, the Hamiltonian of the on-resonance driving fields
is given by
\begin{equation}
H_{I}=\Omega S_{x}=\sqrt{2}\Omega\left(\ket{0}\bra{B}+\ket{B}\bra{0}\right).
\end{equation}
Its eigenstates and eigenvalues are given by $\left\{ \ket+=\frac{1}{\sqrt{2}}\left(\ket B+\ket0\right),\ket D,\ket-=\frac{1}{\sqrt{2}}\left(\ket B+\ket0\right)\right\} $
and $\left\{ \frac{\Omega}{\sqrt{2}},0,-\frac{\Omega}{\sqrt{2}}\right\} $
respectively. Note that all three eigenstates are immune to external
noise. In order to construct an effective $S_{x}^{2}$ (or $S_{y}^{2}$)
driving Hamiltonian of these dressed states, we first need to construct
the couplings $ $ $\ket{-}\bra{+}+\ket{+}\bra{-}$ and $\ket{D}\bra{+}+\ket{+}\bra{D}$
as building blocks, and then use these for the construction of two
off-resonance $\Lambda$ systems, as in the single-drive scheme (see
Fig. $\ref{Fig1}$). By adjusting the phases of the driving fields,
which correspond to the $\ket{-1}\leftrightarrow\ket{0}$ and $\ket{+1}\leftrightarrow\ket{0}$$ $
transitions, the coupling $ $$i\left(\ket{-1}\bra{0}+\ket{+1}\bra{0}\right)+h.c.$
can be constructed. Moving to the dressed states basis, this results
in a $i\left(\ket{-}\bra{+}-\ket{+}\bra{-}\right)$ coupling. Similarly,
by adjusting the phases, the $\ket{D}\bra{0}+\ket{0}\bra{D}$ coupling
is achieved, and adding a phased-matched $S_{z}$ term results in the
desired $\ket{D}\bra{+}+\ket{+}\bra{D}$ coupling. Hence, an effective
$S_{x}^{2}$ Hamiltonian for the dressed states can now be obtained.
Alternatively, it can be shown that the effective Hamiltonian, which
in the bare states basis is given by

\begin{eqnarray}
H_{eff}&=&\frac{\Omega_{2}^{2}}{\Delta}\biggl( S_{z}^{2}\cos^{2}(\frac{\Omega}{\sqrt{2}}t)
S_{y}^{2}\sin^{2}(\frac{\Omega}{\sqrt{2}}t) \nonumber\\
& + & \frac{\sin(\sqrt{2}\Omega t)}{2\sqrt{2}}(S_{y}^{-1}-S_{y}^{+1})\biggr),
\end{eqnarray}
where $S_{y}^{-1}=-i\ket{-1}\bra{0}+h.c.$ and $S_{y}^{+1}=i\ket{+1}\bra{0}+h.c.$,
results in $H_{II}=\frac{\Omega_{2}^{2}}{\Delta}S_{x}^{2}$ in the
dressed states basis, when moving to the dressed states basis and
to the IP with respect to $H_{I}=\Omega S_{z}$. $H_{eff}$ can be
constructed with off-resonance driving fields, similar to the single-drive
construction.

\section{}
\begin{figure}[h]
\includegraphics[width=0.48\textwidth]{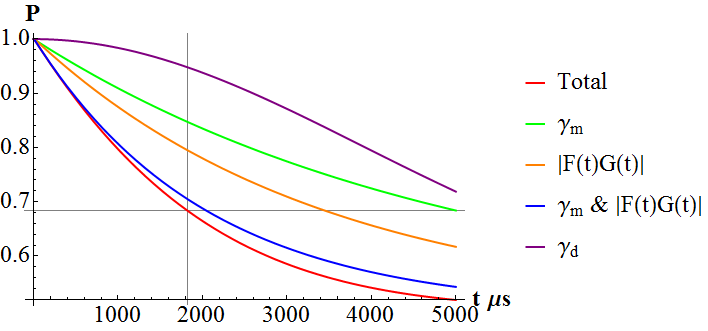}


\protect\caption{\textbf{$P$ as function of $t$.}
$\gamma_{m}\approx200$ Hz - first order dephaing rate due to magnetic noise (green), $|F(t)G(t)|$ - second order dephaing due to magnetic noise (orange), and $\gamma_{d}=182$ Hz - dephasing rate due to noise in the driving fields (purple). The total probability to remain in the initial state (red) is given by $P=\frac{1+|F(t)G(t)|e^{-\gamma_{m}t}e^{-(\gamma_{d}t)^2}}{2}$ (See Eq. \ref{Ptotal}). Gridlines at $P=\frac{1+1/e}{2}$ and $t=1820\:\mu$s.}

\label{T2factors}
\end{figure}

In Fig. \ref{T2factors}  we show the effect of the different sources of noise on the coherence time, which together result in $T_{2} = 1820\:\mu$s.

\section{}
Here we give the values of the parameters used in the estimation of $T_{2}$ in the case of $T_{2}^{*}=3$  $\mu$s. We assume that in all cases we can find a robust point such that  $\gamma_{d}=285$ Hz (compared to $\gamma_{d}=182$ Hz of the simulation).\\

\begin{center}
    \begin{tabular}{ | l | l | l | l |}
    \hline
    $g\mu_{B}B$  (GHz)& $\Omega $ (MHz)& $\Delta_{1} $ (MHz)& $\gamma_{d}$ (Hz)\\ \hline
    10  &  60  & 300  & 285     \\ \hline
    20  &  60  & 300  & 285     \\ \hline
    30  &  75  & 400  & 285     \\ \hline
    40  &  85 & 450  & 285     \\ \hline
    50  &  100  & 500  & 285     \\ \hline
    \end{tabular}
\end{center}


\begin{references}
\bibitem{James1} Note that since we do not make the RWA, the assumption of a narrow band of frequencies, $|\omega_{N}-\omega_{1}|<< \omega_1$, made in Ref. \cite{James} does not hold in this case. This, however, does not effect the validity of the calculation.
\bibitem{James2} The energy shifts due to the effective coupling between the $\ket{B}$ and $\ket{D}$ states are $\sim \frac{\Omega^2 \Delta}{(g \mu_{B}B)^2}$, which is $\sim$ 2 orders of magnitude smaller than the leading contributions of the counter-rotating terms of $\sim \frac{\Omega^2}{g \mu_{B}B}$. We therefore neglect these terms in the calculation of the energy shifts.
\bibitem{DrivingNoise} Under a driving fluctuation, $\Omega\rightarrow\Omega+\delta\Omega$, there is always a point of the parameters such that the energy fluctuations of the $\ket{B}$ and $\ket{0}$ states are equal, $\delta E_{B} = \delta E_{0}$, in the first order in $\delta \Omega$. We assume that we can find a set of values of the parameters which is close enough to this point.
\end{references}
\end{document}